\documentstyle[12pt,epsf]{article}

\newcommand{\AmS}{{\protect\the\textfont2
  A\kern-.1667em\lower.5ex\hbox{M}\kern-.125emS}}
\newcommand{\beq}{\begin{equation}}
\newcommand{\eeq}{\end{equation}}

\title{\hfill {\normalsize KUNS-1561 }\\
Protoneutron stars with kaon condensate}

\author{T. Tatsumi \thanks{This work was supported in 
                part by the Japanese Grant-in-Aid for
                Scientific Research Fund of the Ministry of Education, 
                Science, Sports and Culture (08640369). }
        and 
        M. Yasuhira \\
        Department of Physics, 
        Kyoto University, Kyoto 606-8502, Japan}%

\begin{document}
\maketitle

\begin{abstract}
A new formulation is presented to treat fluctuations around the
kaon condensate. Equation of state (EOS) is given for isothermal and
isentropic cases in the heavy-baryon-limit (HBL). The coexistent 
phase appears in the latter case. The mass-radius relation is given
for protoneutron stars and the possibility of the delayed collapse is
discussed.
\end{abstract}

\section{INTRODUCTION}

There have been extensively studied for years about kaon condensation
and its implications on neutron stars at low temperature \cite{lee}.
In 1994 Brown and Bethe proposed the low-mass black hole (BH) scenario,
based on the large softening of EOS due to
kaon condensation \cite{bro}. It is
produced as a consequence of the {\it delayed collapse} 
from a protoneutron star, different from the usual BH formation.
This scenario should be very attractive in the light of recent 
observations on the mass of neutron stars, SN1987A or future
observation of neutrinos associated with supernova explosions. Some
numerical simulations based on the general relativity have been already 
performed for the delayed collapse \cite{bau,pon}. In ref.\cite{bau} they
studied the delayed collapse by using the EOS of kaon condensed phase
at $T=0$, 
though temperature is very much increased there. Also
neutrino trapping is another important factor to be considered for 
protoneutron stars.

There have been some attempts to treat thermal fluctuations in
relation to kaon condensation \cite{pra}, but there seems to be no
successful theory on the basis of chiral symmetry.
Recently we have proposed a formalism to treat this problem \cite{tat}, 
in relation to protoneutron stars. Here we briefly explain how our formalism
gives the thermodynamic potential, and show some results about EOS and 
structure of protoneutron stars.

\section{THERMODYNAMIC POTENTIAL}

The kaon condensed state can be decribed as a chiral-rotated one from
the meson vacuum \cite{tatb}; actually we can discuss kaon
condensation in almost model-independent way within the mean-field
approximation. However, if we intend to study the phenomenon further 
by taking into account  the effect of fluctuations, it is useful to
invoke the effective Lagrangian like the nonlinear sigma model.

\subsection{Path integral}

We start with the partition function $Z_{chiral}$ for the nonlinear
sigma model ${\cal L}_{chiral}$,
\beq
Z_{chiral}=N\int [dU][dB][d\bar B] \exp[S_{chiral}^{eff}],
\label{aa}
\eeq
with the effective action,
\beq
S_{chiral}^{eff}=\int_0^\beta d\tau\int d^3x
\left[{\cal L}_{chiral}(U, B)+\delta{\cal L}(U, B)\right],
\eeq
where $\delta{\cal L}(U, B)$ is the newly-appeared symmetry-breaking
(SB) term due to the introduction of chemical potentials \cite{tat}. In
evaluating the integral (\ref{aa}), we introduce the local coordinate
around the condensate on the chiral manifold, 
which is equivalent with the following parametrization \cite{tat},
\beq
U\equiv \xi^2=\zeta U_f\zeta(\xi=\zeta U_f^{1/2} u^\dagger=u U_f^{1/2}\zeta),
\quad \zeta=\exp(\sqrt{2}i\langle M\rangle/f),
\eeq
where $\langle M\rangle$ is the condensate, 
$\langle M\rangle=V_+\langle K^+\rangle+V_-\langle K^-\rangle$, with 
$K^{\pm}=(\phi_4\pm i\phi_5)/\sqrt{2}$,
$\theta^2\equiv 2K^+K^-/f^2$ and $V_\pm=F_4\pm iF_5$, while $U_f =\exp[2iF_a\phi_a/f]$
means the fluctuation field.
Accordingly defining a new baryon field $B'$ by way of
$
 B'=u^\dagger B u,
$
we can see that 
\begin{eqnarray}
{\cal L}_{chiral}(U, B)={\cal L}_0(U, B)+{\cal L}_{SB}(U, B)&\longrightarrow&
{\cal L}_0(U_f, B')+{\cal L}_{SB}(\zeta U_f\zeta,u B'
u^\dagger)\nonumber\\
\delta{\cal L}(U, B)&\longrightarrow&\delta{\cal L}(\zeta U_f\zeta,u B'
u^\dagger)
\end{eqnarray}
Thus all the dynamics of kaons and baryons in the condensed
phase are completely prescribed by the non-invariant terms 
${\cal L}_{SB}, \delta{\cal L}$ under
chiral transformation; it is to be noted that the meson mass is
included in ${\cal L}_{SB}$ and the Tomozawa-Weinberg term is in 
$\delta{\cal L}$. 

\subsection{Dispersion relation}

The effective action for the kaon-nucleon sector can be represented as 
\beq
S^{eff}_{chiral}=S_c+S_K+S_N+S_{int},
\eeq
where $S_c$ is the previous classical-kaon  action \cite{lee}
and $S_N$ the nucleon action discarded in HBL. 
The sum  $S_K+S_{int}$ gives the effective kaon action,
\beq
S_K^{eff}=-\frac{1}{2}\sum_{n, {\bf p}}(\phi_4(-n,-{\bf p}), \phi_5(-n,-{\bf
p}))
D^{eff}\left(
\begin{array}{c}
\phi_4(n,{\bf p})\\
\phi_5(n,{\bf p})
\end{array}
\right) +...,
\eeq
with the inverse thermal Green function $D^{eff}$.
Looking for the zeros in $D^{eff}$, we find two solutions $E_\pm$; 
$E_-$ corresponds to the Goldstone mode and exhibits the Bogoliubov spectrum,
\beq
E_-^2\sim \frac{c_3}{2C^2}p^2+\frac{p^4}{4C^2}+...
\eeq
where $C$ means an effective mass for kaons and $c_3$ the product of
the charge density and the $KK$ scattering length \cite{tat}.
We shall see the
importance of the thermal kaon loops due to the Goldstone mode.
The origin of this Goldstone mode is easily understood by observing
that 
the kaon-condensed state is no longer invariant with respect to 
the $V$-spin rotation, while the effective
Lagrangian is still inavariant. 
In other words, we can schematically say 
that the newly-appeared SB term $\delta{\cal L}$ completely cancels  
the original SB term ${\cal L}_{SB}$, and gives rise to a {\it new} 
spontaneous symmetry breaking (SSB) instead.

\section{RESULTS}

First we show EOS thus obtained for the isothermal and isentropic
cases (Fig.~1). 
\begin{figure}[htb]
\epsfxsize=1.0\textwidth
\epsffile{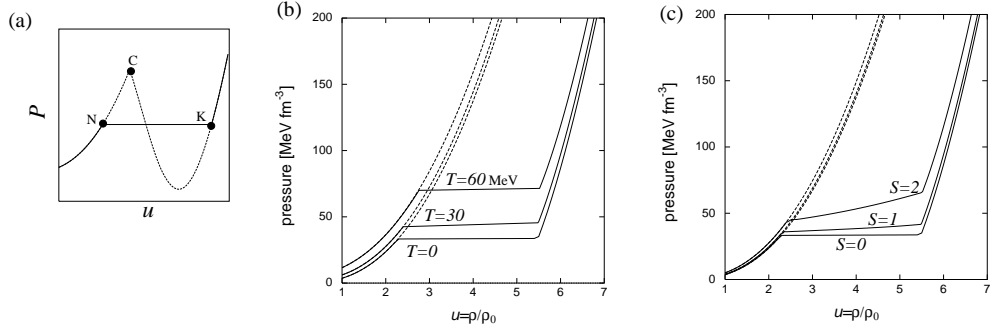}
\caption{Sketch of EOS and graphical construction for the equilibrium
curve in the isothermal case (a). 
EOS for the isothermal case (b) and the isentropic case (c).}
\label{fig:largenenough}
\end{figure}
Since it exhibits the first-order phase transition (FOPT), we prescribe,
for simplicity, the Maxwell construction  
by connecting the equal-pressure densities for the 
normal $(N)$ and condensed $(K)$ states. This means that there is no
coexistent phase for the isothermal matter. On the other hand the
coexistent 
phase appears in the isentropic matter due to the variation of 
temperature density by density.
The magnitude of FOPT is so large that we shall see the existence of
the gravitationally unstable region in the branch of neutron stars
(see Fig. 2(b)).
The effect of thermal fluctuations should be self-evident there.  

For protoneutron stars, the isentropic situation should be more
relevant. In Fig.2(a) we present the temperature profile inside for
protoneutron-star matter
the entropy $s=1, 2$. It is a monotonically increasing function with
respect to density, and takes the maximum value of several tens MeV at 
the center. The difference from no kaon case is rather small. 
In Fig.2(b) we depict the mass-radius relation for protoneutron stars with 
isentropic structure. The branch of protoneutron stars is clearly separated
into two parts by the gravitationally unstable region. We can see that 
thermal effects are insufficient to support higher mass; on the
contrary, the maximum mass is a little bit reduced at $T\neq 0$.
\begin{figure}[htb]
 \begin{minipage}{0.49\textwidth}
  \epsfsize=0.95\textwidth
  \epsffile{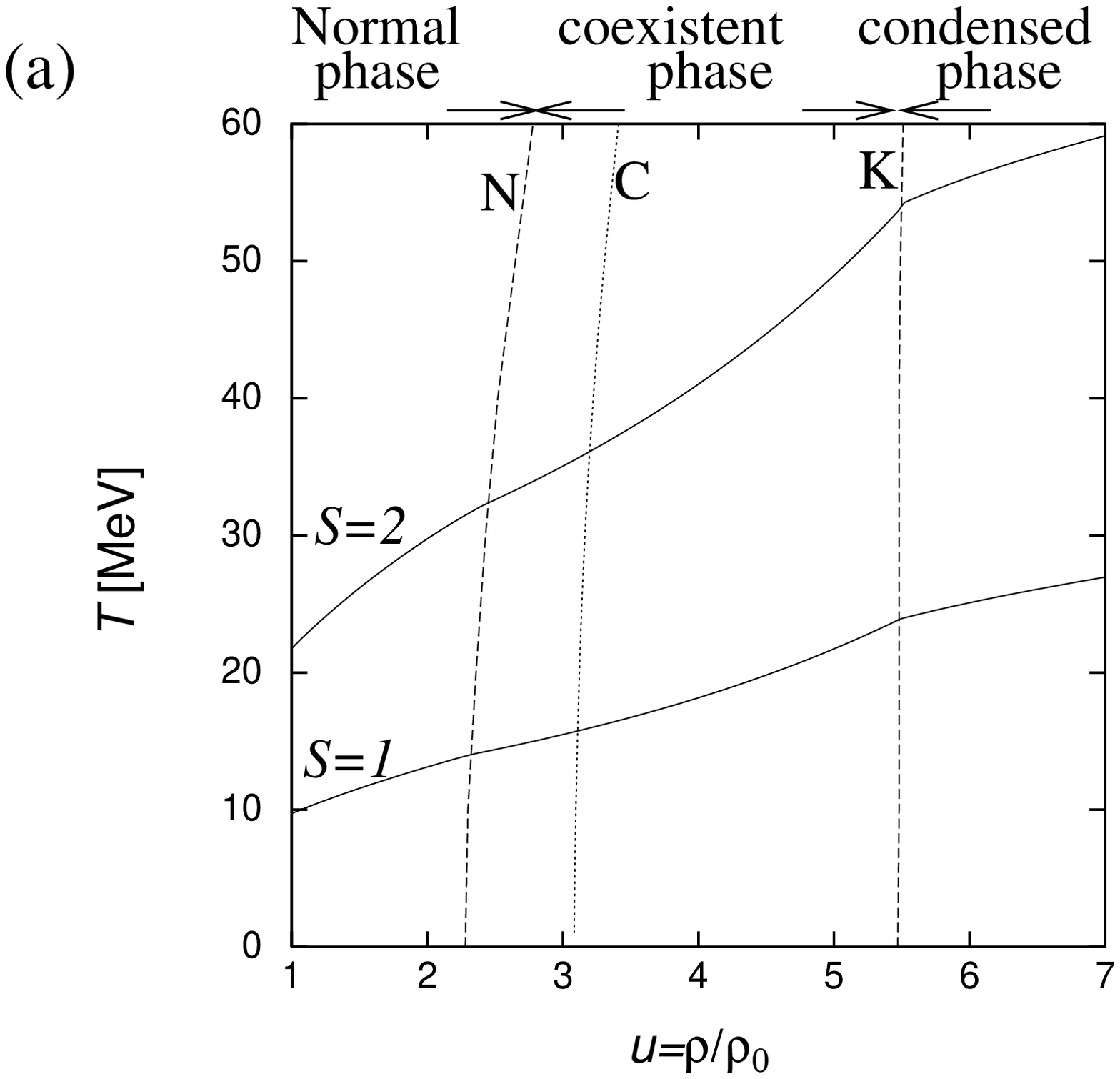}
 \end{minipage}%
 \hfill~%
 \begin{minipage}{0.49\textwidth}
  \epsfsize=0.95\textwidth
  \epsffile{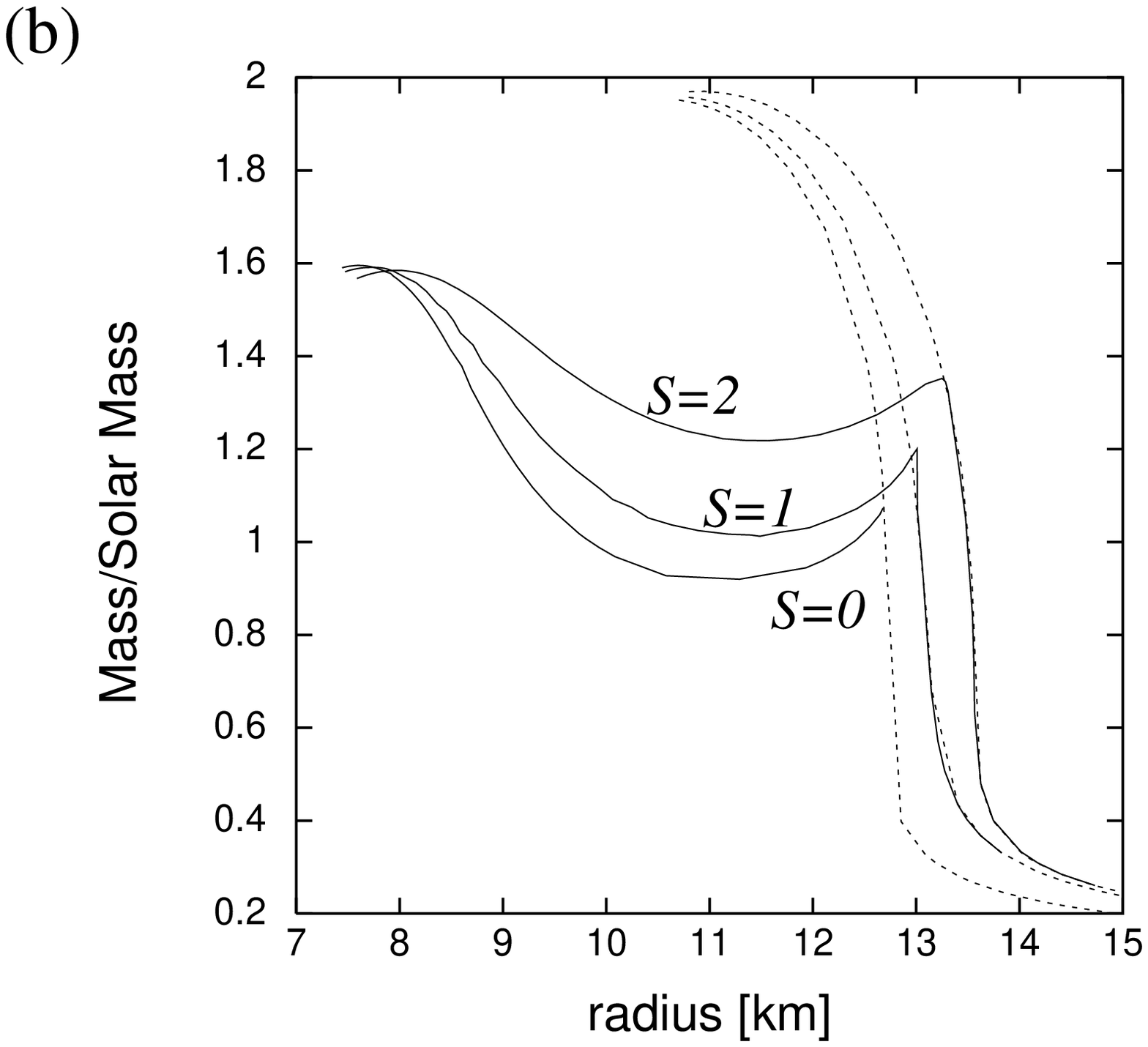}
 \end{minipage}%
\caption{(a)Temperature profile of the kaon-condensed matter.
(b)Mass-radius relation for protoneutron stars (solid
lines). Positive slope region means the gravitationally unstable
branch. The symbols $N, C, K$ correspond to those in
Fig. 1(a). }
\label{fig:toosmall}
\end{figure}

\section{SUMMARY AND CONCLUDING REMARKS}

A systematic formulation to include fluctuations around the condensate 
is presented by introducing the local coordinate on the chiral
manifold. This procedure makes the aspect of chiral rotation prominant 
for the kaon condensation. Using this we obtained the dispersion
relation for the kaonic mode; one is the Goldstone mode as a
consequence of the
SSB of the $V$-spin symmetry, while the other is a very massive mode.

The EOS is obtained, in the HBL, for the isothermal and isentropic 
cases, where the role of thermal kaons should 
be noted. The EOS exhibits FOPT and it might be interesting 
to see the appearance of the coexistent
phase in the isentropic case.
On the other hand, the temperature profile is little changed from 
that in no kaon matter. 
These results may be relevant for the delayed collapse during
the initial cooling era, where neutrinos are no longer trapped. Hence
it is interesting to observe how these results affect the collapsing
process and the profile of the neutrino liminocity by dynamical
simulations.

The maximum mass of protoneutron stars is around $1.6M_\odot$ 
and is not larger than that of cold
neutron stars in this calculation, which suggests more elaborate
study is needed to include nucleon dynamics and the neutrino trapping for
observing the delayed collapse.


\begin{thebibliography}{9}
\bibitem{lee} For a recent review, 
C.-H. Lee, Phys. Reports {\bf 275} (1996) 197.
\bibitem{bro} G.E. Brown and H.A. Bethe, Astrophys. J. {\bf 423}
(1994) 659.
\bibitem{bau} T.W. Baumgarte, S.L. Shapiro and S. Teukolsky,
Astrophys. J. {\bf 458} (1996) 680.
\bibitem{pon} J.A. Pons et al, astro-ph/9807040.
\bibitem{pra} M. Prakash et al, Phys. Reports {\bf 280} (1997) 1.\\
V. Thorsson and P.J. Ellis, Phys. Rev. {\bf D55} (1997)
5177.
\bibitem{tat} T. Tatsumi and M. Yasuhira, Phys. Lett. {\bf B441}
(1998) 9; nucl-th/9811067.
\bibitem{tatb} As a review, T. Tatsumi, Prog. Theor. Phys. Suppl. {\bf
120} (1995) 111, and references cited therein.
\end{thebibliography}
\end{document}